\documentclass[runningheads]{llncs}

\usepackage{hanging}
\usepackage{tabularx}
\newcounter{protocol}

\usepackage{color, colortbl}
\usepackage{nicefrac}
\usepackage{siunitx}
\usepackage{array,framed}
\usepackage{booktabs}
\usepackage{
  color,
  float,
  epsfig,
  wrapfig,
  graphics,
  graphicx,
  subcaption
}
\usepackage{textcomp,amssymb}
\usepackage{setspace}
\usepackage{latexsym,fancyhdr,url}
\usepackage{enumerate}
\usepackage[ruled,vlined]{algorithm2e}
\usepackage{algpseudocode}
\usepackage{graphics}
\usepackage{xparse} % argument parsing -- \edist
\usepackage{xspace}
\usepackage{multirow}
\usepackage{csvsimple}
\usepackage{balance}
\usepackage{
  tikz,
  pgfplots,
  pgfplotstable
}
\usepackage{hyperref}

\usetikzlibrary{
  shapes.geometric,
  arrows,
  external,
  pgfplots.groupplots,
  matrix
}

\pgfplotsset{compat=1.9}
\usepackage{mathtools,}

\DeclareMathAlphabet{\mathcal}{OMS}{cmsy}{m}{n}
\DeclareGraphicsExtensions{%
    .png,.PNG,%
    .pdf,.PDF,%
    .jpg,.mps,.jpeg,.jbig2,.jb2,.JPG,.JPEG,.JBIG2,.JB2}

\begin{document}

\fancyhead{}
\def\thetitle{Securing Password Authentication for Web-based Applications}
\title{\thetitle}

\author{Teik Guan Tan \and Pawel Szalachowski \and Jianying Zhou}

\institute{Singapore University of Technology and Design}

\maketitle

\begin{abstract}
The use of passwords and the need to protect passwords are not going away. The majority of websites that require authentication continue to support password authentication. Even high-security applications such as Internet Banking portals, which deploy 2-factor authentication, rely on password authentication as one of the authentication factors. However phishing attacks continue to plague password-based authentication despite aggressive efforts in detection and takedown as well as comprehensive user awareness and training programs. There is currently no foolproof mechanism even for security-conscious websites to prevent users from being directed to fraudulent websites and having their passwords phished.

\hspace{1.5em} In this paper, we apply a threat analysis on the web password login process, and uncover a design vulnerability in the HTML \texttt{<input type="password">} field. This vulnerability can be exploited for phishing attacks as the web authentication process is not end-to-end secured from each input password field to the web server.
%rather than from the level of the HTML page. In designing our solution, 
We identify four properties that encapsulate the requirements to stop web-based password phishing, and propose a secure protocol to be used with a new credential field that complies with the four properties. We further analyze the proposed protocol through an abuse-case evaluation, discuss various deployment issues, and also perform a test implementation to understand its data and execution overheads.

\end{abstract}

%%
%% Keywords. The author(s) should pick words that accurately describe
%% the work being presented. Separate the keywords with commas.

% Section I
\section{Introduction}
Passwords continue to be popularly used as one (and sometimes the only) method of remote user authentication. It represents the “something-you-know” factor of authentication to allow the other communicating party to verify the identity of the incoming user. This is despite the high levels of phishing and password compromises that organizations face on a daily basis. In the Google 2017 report by Thomas et. al. \cite{thomas2017data}, the authors identified 788k potential victims of keyloggers, 12.4m potential victims of phishing kits and 1.9b user passwords traded on the black market forums over the course of one year. 

Existing best practice recommendations \cite{kskfn2001and,egelman2008you,herzberg2004trustbar}, plus efforts by the community through aggressive detection \cite{marchal2016know} and takedowns \cite{hong2012state}, coupled with investments into user awareness and training have yielded positive but incomplete outcomes. In SonicWall’s 2020 report \cite{sonic2020sonicwall}, phishing volume in 2019 was in its $3^{rd}$ straight year of decline but warns of a growing area related to embedded phishing Uniform Resource Locators (URLs) or web addresses in PDFs and Microsoft Office documents. However, The Anti-Phishing Working Group’s Q2 2020 report \cite{apwg2020phishing} showed that the average number of phishing sites experienced a resurgence back to 50,000 sites globally, after a drop to 40,000 sites back in 2019. Overall email phishing campaigns have dropped from a high of 90,000 in 2017 \cite{apwg2017phishing}  to 45,000 in 2020 \cite{apwg2020phishing}. But APWG warns of the rise in Software-as-a-Service (SaaS), Webmail, and social media sites being targeted for phishing. While the number of phishing attacks may have dropped overall, the value of compromise resulting from such attacks has not waned. The Business e-mail compromise (BEC) metric measures the losses an organization faces due to business email accounts being stolen (typically through a phishing attempt). Also known as the “CXO attack”, the hacker targets a person of authority within an organization to first steal his/her login credentials. Once obtained, the hacker then launches a series of social engineering emails to make the organization carry out activities such as buying gift cards or making payments to bogus accounts. In September 2019, the FBI issued a public service announcement \cite{fbi2019business} warning of the rise in BEC scams which the Internet Crime Complaint Center has received victim complaints with an exposed dollar loss of over \$26 billion for the period between Jun 2016 to July 2019. The Financial Crimes Enforcement Network reported in 2019 \cite{fincen2019updated} that BEC scams continue to generate \$300 million per month in 2018. Recommendations to encourage organizations to move towards multi-factor authentication only prove the point that the current journey to overcome password phishing seems un-ending. 

In this paper, we adopt a different approach to analyze the source of password compromises --- the web browser, and propose mechanisms to remove the possibility of phishing attacks conducted through the web browser. We believe that the proposed mechanisms can be extended to all other forms of password authentication. Our contributions are:
\begin{itemize}
\item Threat modeling and analysis of web password login process which uncovered the design vulnerability of the input password field.
\item Identification of four protocol properties that encapsulate the requirements to stop password phishing for any web login system.
\item Design and validation of a proposed security protocol that complies with the four identified properties.
\end{itemize}

We start with a background of web-based password authentication usage in Section \ref{Section_background}. In Section \ref{Section_threat}, we build the threat model and identify the required properties needed to prevent the disclosure of passwords and misuse of credentials. In Section \ref{Section_related}, we evaluate related works and match them against the identified properties. In Section \ref{Section_protocol}, we construct a security protocol that can meet the properties. In Section \ref{Section_analysis}, we analyze the security and deployment issues of the proposed protocol and benchmark a test implementation of the protocol against existing authentication algorithms. Section \ref{Section_conclusion} concludes the paper.

\section{Background}\label{Section_background}
Banks are one of the largest victims of web-based password phishing due to their customer digitization efforts. The Federal Financial Institutions Examination Council, European Union, and many other regulators have all pushed for standards and regulations \cite{ffiec2005authentication,eu2015directive} in strong customer authentication (SCA), requiring banks to strengthen the authentication process before allowing customers to perform sensitive transactions such as funds transfer and bill payments. SCA regulations mandate banks to either perform risk-based authentication scoring or to supplement the password-based user authentication with a $2^{nd}$-factor token. Risk-based authentication scoring involves collecting specific attributes about the user’s login profile, ranging from IP address, time-of-access, to unique browser and machine attributes, in an attempt to decide if the user is indeed whom he/she claims to be (in addition to providing a password). While it sounds technologically advanced, the effectiveness of risk-based authentication is neither complete nor fool-proof. False-positive and false-negative cases are still high and tend to frustrate valid users and administrators. On the other hand, implementing a $2^{nd}$-factor authentication token has come across as a more effective means to fight phishing. This token can be in the form of a hardware dongle or smartphone that can generate a time-based unique one-time pin (OTP) or using the mobile phone to receive a similar OTP via Short Messaging Service (SMS). As these OTPs are time-bounded, they reduce the time-window opportunity for an attacker to phish for the credentials and then impersonate as the user. But these measures come at a significant cost to organizations, which means that majority of websites (including webmail, enterprise SaaS, and consumer banking sites) still do not enforce 2-factor authentication. Worse still, a poorly implemented $2^{nd}$-factor authentication token without proper precautions in place to protect the $1^{st}$-factor password authentication essentially becomes a weak one-factor authentication system and results in compromised accounts \cite{amnesty2018when} as seen in some cryptocurrency account takeovers \cite{kugler2019the}.

Another area of concern is low password entropy. For account recovery, most mechanisms \cite{reeder2011password} employed by websites allow a user who has lost his/her credentials to enter a secondary password (knowledge-based authentication) or fulfill an out-of-band challenge by entering an OTP. In the survey done by Raponi and Di Pietro \cite{raponi2018spark}, they analyzed 1,000 websites from five European countries and showed that out of the 278 sites that required authentication, the majority of the sites support password-only authentication and they could easily compromise 44\% of their password recovery mechanisms. The problem is more pronounced if the input password field is used for transmission of sensitive information such as the 3-digit Card Verification Value (CVV) for performing credit card payments online. The issue is that these secrets cannot be subject to strong password rules (such as length and characteristics of passwords, use of special characters, etc) and this translates to more points of vulnerabilities for attackers to exploit. 

FIDO (www.fidoalliance.org) is an initiative created as an alliance of vendors (including Amazon, Apple, Facebook, Google, Microsoft, banks, governments, etc) to reduce the use of passwords and improve authentication with hardware dongles or biometric verification. Although FIDO-based authentication is supported on the major platforms, overall FIDO usage is limited (rumored to be in the low millions) which is a drop in the ocean as compared to the billions of users who continue to use passwords to login to their social media and emails. When we examine Facebook's and Gmail's implementation of FIDO, both platforms adopt the U2F standard \cite{dirk2017universal} which still requires the user to enter the password, in addition to presenting the FIDO dongle, for strong authentication.

So the use of passwords and the need to secure passwords are not going away. For many online businesses and websites, password-only authentication systems remain the cheapest to implement and easiest to scale to millions of users, as compared with risk-based, phone-based, token-based or biometric-based authentication. Most websites and cloud SaaS offerings come with a default password authentication setup which can be quickly configured and deployed. While it does not perform bank or government-grade encryption, and only relies on Transport Layer Security (TLS) for session-level encryption, it is functional and commonly used. Even for websites that deploy 2-factor user authentication, such as Gmail and secure Internet banking sites, they continue to rely on password as one factor of authentication. And the reality is that these security-conscious websites do not have a foolproof mechanism to prevent their users from clicking on fake URLs and having their passwords phished by fraudulent websites.

\section{Threat Modelling and Assessment}\label{Section_threat}
At the browser, password entry is handled using the
\texttt{<input type=”password” …>} field as defined in the Hyper-text Markup Language (HTML) standard \cite{faulkner2017html}. Control over the input password field is managed by the website providing the HTML page, where parameters such as maximum and minimum length, regex pattern, and physical size of the field can be defined. Once the password is input by the user or auto-completed by a password manager running on the user’s device, this password is then packaged within an HTML form and submitted to the web server for processing. In most of the phishing cases, users are fooled into clicking on a spoofed URL and are presented with a webpage pretending to be the intended website \cite{hong2012state,thomas2017data}. Users are then inadvertently tricked into entering their passwords into the fake input password field which ends up being read in the clear by the attacker web server rather than by the valid organization. The browser in this case runs on the user’s own device and is not compromised. 

\subsection{Password-Login Use Cases}\label{Section_usecase}
We define a remote web-based password authentication system as a set of one browser-end function, $Derive()$, and two server-end functions $Verify()$ and $Store()$, to perform the following use-cases:
\begin{enumerate}
\item \textit{Password Enrolment}. This is to onboard a user who is new to the system or has forgotten his/her password. In this case, the user is prompted to enter a self-chosen password of sufficient length and character mix, and this chosen password is stored in a derived form for subsequent verification purposes.
\begin{itemize}	
\item In the browser front-end, we define the function $Derive()$ where
\begin{equation} \label{equation_derive}
Derive(UserID,Password) \rightarrow Cred 
\end{equation}
transforms the input password into a credential format that the web server will receive.
\item In the server back-end, we define the function $Store(Cred) \rightarrow output$ which will convert the received credentials into an output for long-term storage. The credential store $CredDB$ will be updated to $CredDB_{new}$. Password enrolment is defined as:
\begin{equation}\label{equation_store}
CredDB_{new} = CredDB \cup Store(Cred)
\end{equation}

\end{itemize}
\item \textit{Password Verification}. This is the most frequently used use-case where the server has to authenticate a previously enrolled user who presents his/her previously chosen password. 
\begin{itemize}
\item In the browser front-end, the same $Derive()$ function (\ref{equation_derive}) is used to transform the password into credentials for server verification.
\item In the server back-end, we define the function $Verify()$ which will return true if the received credentials correspond to a previously enrolled password. Password Verification is defined as:
\begin{equation}\label{equation_verify}
Verify(Cred) \rightarrow true \Rightarrow Store(Cred) \in CredDB
\end{equation}
\end{itemize}
\item \textit{Password Change}. This is for the user to replace his/her chosen password with a new password either on a regular or ad-hoc basis. The change password process requires the user to present both the existing password as well as a newly chosen password, and the operation essentially is the combination of the verification of the existing password and the enrolment of the newly chosen password.
\begin{itemize}
\item In the browser front-end, we apply the same $Derive()$ function (\ref{equation_derive}) on both the existing and new password, obtaining $Cred$ and $Cred_{new}$ to support the change password process.
\item In the server back-end, we authenticate the existing password using the function $Verify()$ before updating on the credential store $CredDB$ with the newly chosen credentials. Password Change is defined as:
\begin{equation}
\begin{aligned}
Verify(Cred) &\rightarrow true \Rightarrow \\
CredDB_{new} &= (CredDB \setminus Store(Cred))\ \cup\ Store(Cred_{new})
\end{aligned}
\end{equation}
\end{itemize}
\end{enumerate}

With the use-cases defined, we can next proceed with threat assessment. The objective is to understand why phishing of passwords continue to happen despite best efforts, and what are the specific requirements to be met in order to eliminate password phishing. 

\subsection{Adversary Model}
The adversary that we are dealing with is a phishing attacker with the goal to authenticate successfully to an honest Web Server on behalf of a user (i.e., in particular, the adversary is interested in stealing users' credentials). We assume that the adversary is able to:
\begin{itemize}
    \item Trick the user to visit a phishing web server under the adversary's control.
    \item Read all transmission data that is sent from the user's browser.
    \item Read all data stored in the honest Web Server's credential store.
\end{itemize}

We further assume that the adversary is not interested in denial of service attacks; the browser software running on the user's device is not compromised; the login session is running over a secure channel (like TLS); and the adversary has no write-access to the honest Web Server or credential store.

\subsection{Threat Assessment}

\begin{figure}[htbp]
\begin{center}
    \includegraphics[width=1.0\textwidth]{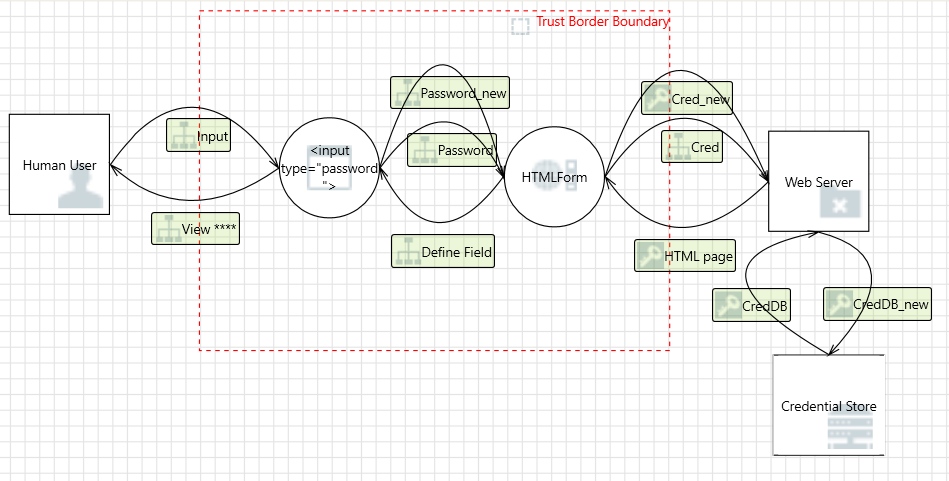}
    \end{center}
\caption{Dataflow diagram of password logins using a web browser.}
\label{Figure_DFD}
\end{figure}

We use the Microsoft Threat Modelling Tool \cite{microsoft2016microsoft} to flesh out the data flow diagram before generating the possible threats. Figure \ref{Figure_DFD} depicts the flow of the password from the user's entry to the browser (depicted by the trust border boundary) before it is submitted to the Web Server for verification. Within the browser trust boundary, the password is received by input password field and collated into the HTML form before it leaves the browser. From the data flow diagram in Figure \ref{Figure_DFD}, the Threat Modelling Tool generated a total of 35 possible threats (excluding denial-of-service threats) which we analyzed individually. We list the threats relevant to password theft and misuse in Table 
\ref{Table_STRIDEthreats}.
\begin{table}[htbp]
\caption{\small
List of relevant threats identified by Microsoft Threat Modelling Tool}
\small
\resizebox{\textwidth}{!}{\begin{tabular}{p{0.7cm}p{3.0cm}p{2.0cm}p{2.0cm}p{7cm}}
\hline
Id & Title	& Category & Interaction & Description \\ \hline
1         & Spoofing of the Web Server External Destination Entity        & Spoofing, Information Disclosure & $Cred, Cred_{new}$ & Web Server may be spoofed by an attacker and this may lead to data being sent to the attacker's target     instead of Web Server. \\ \hline
2         & Data Flow Sniffing  & Information Disclosure & $Input$ & Data flowing across Input may be sniffed by an attacker. Depending on what type of data an attacker can read, it may be used to attack other parts of the system or simply be a disclosure of information leading to compliance violations. \\ \hline
3         & Spoofing the \texttt{<input type="password">} Process      & Spoofing & $Input$ & \texttt{<input type="password">} may be spoofed by an attacker and this may lead to information disclosure by Human User. \\ \hline
4         & Weak Access Control for a Resource  &  Information Disclosure & $CredDB$ & Improper data protection of Credential Store can allow an attacker to read information not intended for disclosure. \\ \hline
5           & Spoofing the Web Server External Entity or Elevation by Changing the Execution Flow in HTMLForm    & Spoofing, \newline Elevation of \newline Privilege & $HTMLForm$ & Web Server may be spoofed by an attacker and this may lead to unauthorized access to HTMLForm. Or an attacker may pass data into HTMLForm in order to change the flow of program execution within HTMLForm to the attacker's choosing. \\ \hline

\end{tabular}}
\label{Table_STRIDEthreats}

\end{table}

One vulnerability pointed out by Threat Id 1 and 5 of Table \ref{Table_STRIDEthreats} is at the design of the input password field in handling the password. At present, the input password field masks off the user input from the display, ensuring that information entered in that field cannot be seen by the user or anyone physically in front of the screen with the user. However, when we examine within the browser, the said field contents are accessible to any malicious components (e.g. Javascript) in the clear within the HTML page. Furthermore, when the input password field is submitted as part of the HTMLForm to the Web Server, the contents of the field are in the clear, allowing the communicating Web Server to gain access to the contents. In order to phish for passwords, an attacker would need to register a domain name and set up a website that closely resembles the real site. Then the user is lured to the attacker's site through spoofing URL links and tricked into entering his/her password credentials into the fake HTML page. Since the fake HTML page will transmit the input password field contents in the clear to the attacker's website, passwords are harvested by such phishing attacks. These passwords can then be replayed as many times as the attacker wants to impersonate as the user. For such a design, the burden is placed on the user to check the URL before entering the password and unfortunately, mitigation efforts to train users to recognize spoofing URLs have only seen limited success \cite{egelman2008you,thomas2017data,gupta2018defending}. 

Another vulnerability highlighted by Threat Id 4 has to do with credential verification. Simple password authentication is vulnerable to replay attacks where the credentials submitted by the web browser can be re-submitted to gain the same level of access. Since these submitted credentials may potentially be captured in the audit or debug logs on the web server, we need to ensure that the submitted credentials cannot be re-usable for another login. The same logic goes for the credentials stored in the credential store. The credential store functions as the “source of truth” for the users’ secret password. The stored credentials are referenced by the Web Server for purposes of verifying against the password presented by the user, but these stored credentials should not be usable as credentials to impersonate the user during password logins.

Finally, the input interaction between user and browser is a point of vulnerability mentioned by Threat Id 2 and 3, where keyloggers and other malware can capture the password as the user enters it into the input password field. Attackers can fool users by creating a text box that mimics an input password field, and trick users into entering their passwords into the text box. Alternatively, the password input by the user may be sniffed by keyloggers. This latter threat happens at the operating system layer, outside the context of the browser, and is probably the best-understood vulnerability. Most, if not all, anti-virus software have built-in detection to deal with keyloggers, and the result is the low (less than 1\%) level of compromise (as seen in Google’s report \cite{thomas2017data}) when compared to phishing attacks and stolen user credentials.

In summarizing the threat findings, it is clear that the current design of the input password field is a source of vulnerability for multiple threats: 
\begin{itemize}
    \item Within the browser, any active code such as Javascript has access to the clear password values contained in the input password field.
    \item Within the web session, any password content that is submitted with the HTMLForm to any Web Server is accessible in the clear.
    \item Outside the web session, any attacker can spoof a fake input password field and possibly fool end-users into entering their password into such a field.
\end{itemize}
And these need to be comprehensively addressed in order to stop phishing.

\subsection{Requirements Considerations}

We start by addressing the design of the input password field. We recommend that the existing functionality of the input password field to be deprecated, and a new input credential field (such as \texttt{<input type="credential"..>}) to be introduced to prevent not just other users from seeing the input, but also any other active components or forms within the HTML page from accessing the raw input. No field that mimics password entry \footnote{Browsers can achieve this by making the look-and-feel of the new input credential field \underline{not possible} to be created by any CSS style sheet or HTML code.} should be allowed in the HTML page. In addition, the password provided by the human user into the new input credential field should never be allowed to be read or submitted in the clear and instead be converted into credentials for the sole purpose of one-time validation by the web server. We consolidate the requirements into the following four properties:
\begin{itemize}
\item	\textbf{P1 - Credential Derive one-way function}. When the credential is read from the input credential field for HTML manipulation or during submission, it is not possible for an attacker to obtain the password from the credential. 
\item \textbf{P2 – Credential Specific one-time use}. When the password is entered into the input credential field, it can only be used at that specific time, from that specific browser, for the specific purpose of authenticating the specific user to the specific web site. Knowledge of the credential should not allow the attacker to derive past credentials or generate future credentials.
\item \textbf{P3 – Credential Store one-way function}. When the credential is submitted to the website and stored in the credential store during password enrolment or password change, it is not possible for an attacker to obtain the credential from the stored credential.
\item \textbf{P4 – Non-interactive input credential field}. When the input credential field receives a password entry from the user, it should be able to perform the $Derive()$ function non-interactively to output the credential that is usable for both enrolment and verification. The credential should not depend on any HTML state or require any protocol exchange with an external system. This is to prevent an attacker from exploiting any possible intermediate states between the password and credential.
\end{itemize}

We compare these four properties against the R1 to R10 password requirements listed by Liao et. al. \cite{liao2006password} and believe that these four provide a more objective means of evaluation, except for R9 mutual authentication which is not a requirement in our case.
When applied back to the identified threats in Table \ref{Table_Mitigation}, we can see that these four properties will comprehensively address all the vulnerabilities.

\begin{table}[htbp]
\caption{Application of properties to threats from Table \ref{Table_STRIDEthreats}}
\small
\resizebox{\textwidth}{!}{\begin{tabular}{p{0.8cm}p{3.1cm}p{12.0cm}}
\hline
Id & Addressed by	& Explanation \\ \hline
1         & P1 \& P2        &The spoofed web server receiving the credentials will not be able to reverse the passwords from the credentials or to use the stolen credentials to authenticate against a different web server   \\ \hline
2         & P2 & The valid web server needs to verify credentials that can be traced to the time, place, and purpose of the password entry. The mere possession of the password which can be replayed should not be accepted.  \\ \hline
3         & Deprecating the input password field \& P1        &The browser must never return the cleartext password from the input credential field to any active components within the HTML   \\ \hline
4         & P3        &Stolen stored credentials will not allow an attacker to guess the credentials from the stored credentials. \\ \hline
5         & P4         &In order to prevent a malicious HTML to manipulate the generation of credentials, the input credential field needs to be able to derive the credentials in a stateless, non-interactive manner. \\ \hline
\end{tabular}}
\label{Table_Mitigation}
\end{table}

\section{Related Works}\label{Section_related}
The threats posed to password-based authentication have also been studied widely in the research community.

Much of the efforts in strengthening password security happen between the browser and the website, as well as how the credentials are stored in the backend. Haller's S/Key \cite{haller1994s}, Lamport's one-time password \cite{lamport1981password} and many other proposals \cite{bonneau2011getting,yang1999password} exist, but all require state information to be retained at the client which violates Property P4. Encrypted key exchange (EKE) was first proposed by Bellovin and Merritt in 1992 \cite{bellovin1992encrypted} where they effectively used a combination of symmetric and asymmetric key cryptography to encode the password in transit to defend against eavesdropping. The EKE protocol, however, is interactive in nature and requires the backend server to possess the same password value as the user in order to perform the authentication. This breaks both Properties P3 and P4. The Secure Remote Password (SRP) \cite{wu1998secure} protocol improves on EKE by not requiring the storage of the plaintext password in the server. However, SRP cannot be applied for password enrolment or change use-cases, and requires the exchange of session random between the client and server which violates the non-interactive requirement in Property P4. On the browser-end, Boyen \cite{boyen2009hpake,boyen2007halting} advocated the need to protect passwords against impersonation servers. This goal is echoed by EKE, SRP and other password-authenticated key exchange schemes \cite{jablon1996strong,bellare2000authenticated,jarecki2018opaque,ieee20091363} which aim to mitigate the exposure of low-entropy user passwords \cite{wu1999real} through conveying as little or zero password information during the password verification, but this protection does not extend to password enrolment. 

An alternative approach is to discard the symmetric nature of passwords and use asymmetric key cryptography for authentication. The Transport Layer Security (TLS) protocol \cite{rescorla2018transport}, formerly known as Secure Socket Layer, includes optional support for client-side authentication as part of the TLS handshake process where the user uses X.509 credentials to sign the server challenge and gets authenticated by the corresponding web server. The setup relies on the use of certification authorities to ascertain the identities of the users and issue X.509 certificates that link the identities of the users to associated public/private key pairs. Users store their private keys either in hardware dongles or in encrypted software files on their devices. The password is relegated to protect access to the user’s private key associated with the X.509 certificate, and does not feature cryptographically in the authentication process. While both the security strengths and weaknesses of TLS client authentication have been well studied \cite{parsovs2014practical}, the complicated certificate management requirements coupled with a more cumbersome user experience has hindered widespread adoption of TLS client authentication as compared to password authentication. An interesting development proposed by Delignat-Lavaud et. al. \cite{delignat2016cinderella} implemented verifiable computation, a family of zero-knowledge proofs, for X.509-based user authentication to improve on the privacy and operating overheads. Other enhancements to TLS include Dietz et. al. \cite{dietz2012origin} proposal to supplement the authentication process by using self-signed client certificates as bearer tokens, as well as efforts to incorporate SRP into TLS \cite{taylor2007using}. The attractive characteristic of using asymmetric key cryptography for authentication is that it is able to meet Properties P3 and P4. The design challenge is how to make the input credential field transform a low-entropy password input into an asymmetric key digital signature.

%One of the largest asymmetric key authentication systems deployed, with approximately 2 to 5 million active users, is the Bitcoin network \cite{nakamoto2008bitcoin}. Users mainly make use of software wallets which store the secp256k1 ECC (Elliptic Curve Cryptography) private key. Private keys are randomly generated or derived and each ECC key pair has a corresponding wallet address which is a double hash (of RIPEMD-160 and SHA2-256) of the ECC public key. No certification authorities are used. During authentication, a user enter the password to unlock the wallet and use the ECC private key to cryptographically sign Bitcoin transactions which can then be verified by the network. The point here is that Bitcoin does not rely on any identity certificates and still could operate a distributed, large-scale asymmetric key authentication system effectively.

The third course of action is to take the user out of the password entry equation and rely on software password managers to input highly complex passwords, maintain different passwords for different accounts, and perform regular password changes, in an effort to strengthen password authentication while reducing the burden on the user. Password managers such as Keepass (www.keepass.info), Lastpass (www.lastpass.com) and Google password manager store and protect the user credentials on behalf of the user, and only require the user to remember a master password to unlock all the credentials. One problem with password managers, besides the obvious concentration risk where all passwords are placed in one basket, is the difficulty for users to tell the difference between good and bad password managers. There are many choices of password managers and not all are equally good \cite{silver2014password}. Good password managers should minimally be able to detect if the HTML page requesting for the password is indeed a valid page or a phishing page, and have to be constantly updated whenever the website layout changes or when hackers improve their tactics. But this is not foolproof \cite{winder2019google,simplicit2018browser}, and there is no guarantee the vendors who provide password managers can keep up with malicious activity. Those that focus on security, such as Keepass which has a bug bounty page and security certifications by multiple government agencies, may not be the most convenient-to-use or may lack some of the features required by different users \cite{fitzpatrick2020password}. From an applicability perspective, password managers do not meet Properties P1 nor P2.

\section{New Password Login Protocol}\label{Section_protocol}
To meet the four properties identified in Section \ref{Section_threat}, we leverage much of previous work done on increasing password entropy, device verification, and asymmetric key authentication to provide a suitable construction. 

\subsection{Design Strategy}\label{section_strategy}

Our design strategy to meet the four principles is as follows:
\begin{itemize}
    \item \textit{Input field-level $Derive()$ operation}. To meet Property P1, the password should automatically be converted by the browser into a one-way credential once it is entered into the input credential field. This takes away the possibility for any malicious code within the HTML page or remote web server from getting access to the clear input password from the user.  
    \item \textit{User, Device, Time, Server, and Session dependent credential}. To meet Property P2, the $Derive()$ function needs to convert the input password into verifiable credentials which cannot be reused, redirected, reversed or offline dictionary attacked:
    \begin{itemize}
        \item To prevent reuse, the credentials need to include a timestamp and session challenge.
        \item To prevent redirection, the credentials need to include the URL from which the browser has loaded the HTML page.
        \item To prevent reversal, we propose the use of Password-Based Key Derivation Function 2 (PBKDF2) \cite{moriarty2017pkcs} as the strong one-way function to transform the password. PBKDF2 is a one-way function that takes in the password and salt as input, and carries out multiple iterations of a deterministic pseudo-random function (PRF) such as SHA2-256 to output a key value. Alternatives to PBKDF2 include Argon2 \cite{biryukov2016argon2} or Halting KDF \cite{boyen2007halting}.
        \item To prevent offline dictionary attacks, the $Derive()$ function needs to be able to protect low-entropy passwords (such as 6-digit pin, or payment card CVV) from being brute-force guessed by the attacker. Our strategy is to include browser authentication in the credentials. We require all browsers to contain a random, secret, and un-exportable key $S_b$ that can be used to sign the credentials. This is similar in purpose to Dietz et. al. \cite{dietz2012origin} origin-bound certificates, albeit at the input credential field level rather than at the HTML level. We propose the use of ECDSA as the signing algorithm with a Trust-on-First-Use policy to remove any need for key management. ECDSA is proven under Brown's generic DSA model \cite{Brown00theexact} to be EUF-CMA \cite{goldwasser1988digital}.
    \end{itemize}
    \item \textit{One-way deterministic $Store()$ function}. To meet Property P3, the $Store()$ function needs to perform a one-way operation to transform the credentials for storage and use in subsequent password verification. We propose to similarly use PBKDF2 for this purpose.
    \item \textit{Common browser-end $Derive()$ function}. To meet Property P4, the input credential field in the browser should perform the same $Derive()$ function regardless whether the input password is to be used for password enrolment, verification or change. Such a function is therefore not browser-state dependent, and does not have intermediate states for the attacker to exploit. To achieve this, the credentials produced from the input password needs to include four parts:
    \begin{itemize}
        \item Proof of knowledge of $password$ that is entered into the input credential field.
        \item Unique $password$ identifier for storage during enrolment or comparison during verification.
        \item Proof that $password$ is entered in a specific browser.
        \item Unique browser identifier for storage during enrolment or comparison during verification.
    \end{itemize}
\end{itemize}

\subsection{Proposed Protocol}\label{section_proposed}

Using the strategy described in Section \ref{section_strategy}, we design the password enrolment, verification and change protocol as a single 2-way exchange of information starting from the Web Server $\rightarrow$ Browser $\rightarrow$ Web Server:
\begin{enumerate}
    \item Web Server: Upon receiving a page request from the browser, the Web Server will return a randomly generated challenge with the HTML page.
    \item Input Credential Field in Browser: Upon the end-user entering the password, the field will call $Derive()$ to convert the password into credentials for HTML form submission.
    \item Web Server: Upon receiving the credentials, 
    \begin{enumerate}[a]
    \item If this is a password enrolment request:
    \begin{itemize}
        \item Call $Store()$ to convert the credentials for storage.
        \item Update the credential store with the user's storage credentials.
    \end{itemize}
    \item If this is a password verification request:
    \begin{itemize}
        \item Call $Verify()$ to ascertain the validity of the credentials.
    \end{itemize}
    \item If this is a password change request, then 2 sets of credentials (existing and new) will be received:
    \begin{itemize}
        \item Call $Verify()$ to ascertain the validity of the existing credentials.
        \item If the check is successful, call $Store()$ to convert the new credentials for storage.
        \item Replace the existing storage credentials in the credential store with the user's new storage credentials.
    \end{itemize}
    \end{enumerate}
\end{enumerate}
We provide the construction of the 3 functions, $Derive()$, $Store()$ and $Verify()$ below. Our construction can be extended by the existing interactive PAKE protocols \cite{wu1998secure,jarecki2018opaque}, as long as they meet the four principles. 

\subsubsection{Browser-end Function $Derive()$}
The function $Derive()$ is run on the browser for each input credential field and takes in the input password to output one-time use credentials that can be used by the website for all three use-cases of password enrolment, verification and change. Our proposed $Derive()$ function is described in Algorithm \ref{algo_derive}:

\SetAlgoNoLine
\begin{algorithm}[H] \label{algo_derive}
  \DontPrintSemicolon
  \KwIn{$UserID, Challenge,Password$}
  \KwOut{$\sigma_p,\sigma_b,V_p,V_b,BrowserTime$}
  $URL = GetBrowserURL();$\;
  $S_p = PBKDF2(Salt=URL||UserID,Password)$;\;
  $\sigma_p = ECDSA.Sign(Challenge,S_p)$;\;
  $V_p = ECDSA.GetPublicKey(S_p)$;\;
  $delete\ S_p$;\; 
  $BrowserTime = GetSystemTime()$;\;  
  $S_b = GetBrowserKey()$; \Comment{to get handle to browser key}\;
  $\sigma_b = ECDSA.Sign(\sigma_p||BrowserTime,S_b)$;\;
  $V_b = ECDSA.GetPublicKey(S_b)$;\;
  \Return $\sigma_p,\sigma_b,V_p,V_b,BrowserTime$
  \caption{Proposed $Derive()$ function}
\end{algorithm}

$Derive()$ uses PBKDF2 (to prevent reversal) where the page $URL$ and $UserID$ are concatenated as the salt value (to prevent redirection) to transform $Password$ into an internal secret key $S_{p}$. It next uses $S_{p}$ to ECDSA sign the $Challenge$ received from the Web Server (to prevent reuse) to generate a signature $\sigma_{p}$ as proof of knowledge of $Password$. $V_p$ is $S_p$'s corresponding public key for use to verify $\sigma_p$. To prevent offline dictionary attacks, $Derive()$ will use the internal browser key $S_b$ to ECDSA sign $\sigma_p$ concatenated with the current time, producing $\sigma_b$. $V_b$ is $S_b$'s corresponding public key for use to verify $\sigma_b$. 

\subsubsection{Server-end Function $Store()$}
The function $Store()$ transforms the received password public key $V_{p}$ and browser public key $V_{b}$ into identifiers for storage to make them more resistant to any rainbow dictionary attacks \cite{oechslin2003making}. Our proposed $Store()$ function is described in Algorithm \ref{algo_store}:

\SetAlgoNoLine
\begin{algorithm}[H] \label{algo_store}
  \DontPrintSemicolon
  \KwIn{$salt,V$}
  \KwOut{$P$}
  $P = PBKDF2(salt,V)$;\;
  \Return $P$
  \caption{Proposed $Store()$ function}
\end{algorithm}

$Store()$ uses PBKDF2 with a salt to derive the public key into an identifier for purposes of storage and subsequent verification. For the password identifier, we propose that $UserID$ be used as the salt. For the browser identifier, we propose that a $null$ value be used as the salt. This is to allow organizations that operate risk analytics systems under the SCA regulations \cite{eu2015directive} to make use of browser identifier as an additional attribute and blacklist browsers that may be used for malicious purposes. 

\subsubsection{Server-end Function $Verify()$}
The function $Verify()$ is a function run by the Web Server to authenticate the remote user. It has to verify the validity of the password ECDSA signature $\sigma_{p}$ and the browser ECDSA signature $\sigma_{b}$. Our proposed $Verify()$ function is described in Algorithm \ref{algo_verify} :

\SetAlgoNoLine
\begin{algorithm}[H] \label{algo_verify}
  \DontPrintSemicolon
  \KwIn{$UserID,Challenge,BrowserTime,\sigma_p,\sigma_b,V_p,V_b$}
  \KwOut{$result \in {true,false}$}
  $ServerTime = GetSystemTime()$;\;  
  \If {$ServerTime - BrowserTime > \delta$} 
  {\Return $false$;}
  $P_p = GetStoredPasswordID(UserID)$;\;
  $P_b = GetStoredBrowserID(UserID)$;\;
  \If {$(P_p \neq Store(UserID,V_p))\ or \ (P_b \neq Store(null,V_b)) $}
    {\Return $false$;}
  $result = false$;\;
  \If {$ECDSA.Verify(Challenge,\sigma_p,V_p) == true$}
  {\If {$ECDSA.Verify(\sigma_p||BrowserTime,V_b) == true$}
  {$result = true$;} }
  \Return $result$
  \caption{Proposed $Verify()$ function}
\end{algorithm}

$Verify()$ performs three checks. Firstly, the time difference between the browser signing the credentials to the server verifying the signature should be within an acceptable $\delta$ time window. Next, the password public key and browser public key should match the identifiers that were previously enrolled and stored. Finally, the password and browser signatures need to validate correctly to fully ascertain that the user did enter the previously-enrolled password in a recognized browser. For practical purposes, $Verify()$ can be modified to allow a user to use multiple browsers, instead of only a single pre-enrolled browser, to login to the same web site. We discuss this under User Experience in Section \ref{section_issues}.

\section{Analysis}\label{Section_analysis}

In this section, we analyze the security, deployability, and implementation issues of the proposed protocol. We study the security of the proposed protocol by putting it through a set of known abuse cases. Various deployment topics are raised and discussed. A test implementation is done to evaluate its data and execution overheads.

\subsection{Abuse cases}
Besides designing our protocol to meet the four properties, we further validate our protocol through known abuse cases. Table \ref{Table_Abuse} lists the evaluation of different attacks on our protocol. Aside from a compromised browser and other system-level attacks which are outside the scope of the model, we are assured that a security-conscious web site will be able to prevent password phishing with our proposed input credential field, provided that the existing input password field is deprecated and the Web Server is running over TLS.

\begin{table}[t!]
\caption{Evaluation of abuse cases on the proposed protocol}
\small
\resizebox{1.0\textwidth}{!}{\begin{tabular}{p{2.0cm}p{4.0cm}p{6.0cm}p{2.3cm}}
\hline
Attack & Abuse case	& Result & Remarks \\ \hline
User visits \newline spoofed URL & Attacker steals credentials as man-in-the-middle to attempt to login to the real URL & $Derive()$ computes credentials which include the spoofed URL. This will fail verification with the real URL. & No compromise. Principles P1 \& P2 are satisfied. \\ \hline
& Attacker tries to offline brute-force guess password using stolen credentials & Attacker is unable to generate $\sigma_p$ for passwords with sufficient entropy. Attacker is unable to generate $\sigma_b$ which requires the browser key $S_b$ \newline \textit{Assumption:} Browser is not compromised. & No compromise. Principle P1 is satisfied.
 \\ \hline
$Time$ on \newline end-user's \newline device is \newline modified & Attacker changes the time on the user’s device to a future date to steal future-dated credentials & $Derive()$ computes credentials which include an incorrect server challenge \newline
\textit{Assumption:} Attacker is unable to predict the server challenge	& No compromise. Principles P2 \& P4 are satisfied. \\ \hline 
Password is \newline sniffed at \newline input & Attacker captures user input using keylogger and uses another device for login & 
Attacker is unable to generate $\sigma_b$ which requires the browser key $S_b$ \newline \textit{Assumption:} Browser is not compromised. & No compromise. Principle P2 is satisfied.
\\ \hline
Compromised Browser &	Attacker installs a trojan browser and has control over the browser operations & Outside the scope of protection  & Require to install anti-virus.
 \\ \hline
Browser memory captured &	Attacker does a system-level memory dump of the client machine to capture intermediate values and Browser key $S_b$ &	Outside the scope of protection &	Require to install anti-virus\\ \hline
Multiple \newline credentials \newline harvested &	Attacker collects as many $V_p$, $\sigma_p$, $V_b$, $\sigma_b$ as possible to guess the password or derived key $S_p$ or generate a valid $\sigma_p$ & Not possible to guess or generate alternate credentials as ECDSA is EUF-CMA \cite{goldwasser1988digital}.  & 	No compromise. Principle P2 is satisfied.\\ \hline
Stolen server credentials &	Attacker collects stored $P_p$ in an attempt to guess the password or generate a valid $\sigma_p$ &	Not possible to guess or generate alternate credentials as ECDSA is EUF-CMA \cite{goldwasser1988digital}. &	No compromise. Principle P3 is satisfied.\\ \hline
Credentials replayed & Attacker collects $\sigma_p, V_p, \sigma_b, V_b$ and replays the credentials for login & The replayed credentials include an incorrect server challenge \newline
\textit{Assumption:} Server challenge is unpredictable and not repeated.
& No compromise. Principle P2 is satisfied.\\ \hline
Credentials replaced during enrolment or password change & Attacker replaces the new credentials $\sigma_p^{new}, V_p^{new}, \sigma_b^{new}, V_b$ in transit with fake credentials& This would require the attacker to compromise the TLS session between the browser and web server. \newline \textit{Recommendation:} web server should be running TLS \cite{rescorla2018transport}, with HTTP Strict Transport Security (HSTS) \cite{hodges2012http} turned on. & No compromise. Principles P2 \& P4 are satisfied.\\ \hline

\end{tabular}}
\label{Table_Abuse}
\end{table}

\subsection{Deployability}\label{section_issues}
We recognize that the proposed input credential field will require significant system changes and testing for the front-end browsers and back-end web servers as well as credential repositories to support this new protocol. But this is not without precedence. One of the biggest changes to web browsing over the past two decades is the widespread use and eventual demise of Java Applets \cite{vice2016applet}. Many organizations such as banks and enterprises used Java Applets to augment the authentication process, and hence removing the support for Java Applets, partially due to the security issues it posed, was met with skepticism. It took several years and the community's efforts to strengthen the Javascript functionality before Java Applet's use was diminished to insignificance. Here, we briefly highlight other deployment issues that may arise due to the proposed protocol. A more in-depth study is required and is left for future research.

\subsubsection{Password Migration}
With the proposed change, existing password repositories and backups will cease to be usable but instead may pose a vulnerability if left exposed. Beyond system changes, all HTML application code  with password login pages have to be updated and web applications are required to handle the migration of backend password credentials from the existing format to the new proposed format.

A 2-step solution is to first introduce the input credential field to exist concurrently with the input password field. This will allow for the gradual migration of passwords which can happen seamlessly upon the users' next login. Once the defined time-window (of possibly 6 to 12 months) is up, the input password field can then be deprecated and ignored by the new version of browsers. We believe this migration can be successfully executed by the top 3 browser companies, namely Google Chrome, Apple Safari, and Mozilla Firefox, as they presently command a combined market share of over 85\% of the worldwide browser market \cite{statcounter2020desktop}.

\subsubsection{User Experience (UX)}
An issue to consider is the user experience when users use different browsers for access to the same website. Users may own multiple devices or may change devices from time to time. They may also log in using publicly-accessible shared browsers from Internet cafes. The range of responses that the web server can perform is wide and varied, depending on the level of security the application requires versus the amount of UX friction that the user is prepared to accept. We list 3 possible scenarios here:
\begin{itemize}
    \item \textit{High Security.} If a sensitive application such as Internet Banking system detects a new browser, the server should issue a step-up authentication challenge, such as sending an OTP via SMS to the user's pre-registered mobile phone, and only allow access to proceed if the OTP challenge is correctly entered. This is to ensure that the user's login credentials were not obtained illegally and that the user is aware of this login taking place. The server should keep a history of 3 to 5 browsers for each user, but exclude browsers that are widely shared (i.e. many users having the same $P_b$).
    \item \textit{Enterprise Security.} For cloud-based SaaS business applications, business webmail, or enterprise Virtual Private Network (VPN) remote access, the server should log any access from an unrecognized browser, and send an email alert to the user whose credentials are used for login. Access can still proceed but the user may be required to acknowledge the email alert within 24 to 48 hours or risk triggering an investigation into irregular access. The server should keep a history of 3 to 5 browsers for each user, and similarly exclude browsers which are widely shared (i.e. many users having the same $P_b$).
    \item \textit{Personal Security.} For personal applications such as email or social media, the server could keep a history of the past 8 to 10 browsers used by each user, and send an email notification (to an alternate recovery email address) whenever a new browser is detected. Advice on what the user should do can be included in the notification. In addition, the server should maintain a blacklist of suspicious browsers that are possibly used for stolen credentials and deny any access from these browsers.
\end{itemize}

\subsubsection{User Privacy}
Since the Web Server stores one or more browser identifiers $P_{b1}, P_{b2}, ...$ associated with each user, it has the data to correlate browsers to different groups of users and use this information to deduce real-world relationships and other possible user associations. This issue is similarly observed by Dietz et. al. \cite{dietz2012origin} who mentioned that the use of client-side browser signing will introduce privacy issues. One possible method to enhance the users' privacy is to use a user-specific salt when calling the $Store()$ function to convert the browser public key $V_b$ into the browser identifier $P_b$, but this will retard the use of blacklists to deter attacks. 

\subsubsection{Password Entry Checks}
The proposed solution requires the input password to be derived at the browser's input field which prevents any possibility of checking for password strength at the server. If the website requires specific password characteristics (e.g. combination of upper/lower case, numeric, etc), it will have to be passed into the input credential field and enforced by the browser.

Conversely, practices to require the user to enter a new password twice (as a consistency check) during password enrolment or password change may be impacted. In our proposed protocol in Section \ref{Section_protocol}, this consistency check can be carried out at the browser by comparing the values of $V_p$ from both fields, but this may not be possible if other algorithms are used.

\subsubsection{RADIUS and other Legacy Authentication}
The biggest impact caused by the input credential field change is likely to be web-based systems that still use legacy password protocols such as Remote Authentication Dial-in Service (RADIUS) \cite{rigney2000remote} in the backend. In such systems, the web server may not perform $Verify()$ function as described in Section \ref{section_proposed}, but instead rely on a separate authentication service to carry out the credential verification. The SSL VPN gateway is an example of such an implementation. During authentication, the VPN gateway receives the userID and password entered by the user and functions as a RADIUS client to transmit the userID and password to a RADIUS server in an ACCESS-Request packet to check for the validity of the credentials, before establishing a TLS encrypted tunnel between the user's machine to the VPN gateway. The RADIUS protocol assumes the RADIUS client to have clear-text knowledge of the password in order to form the ACCESS-request packet, and this will not be possible with the proposed input credential field.

A solution would be to review and update the RADIUS protocol to support a new credential attribute, or to expand the size of the current User-Password attribute from 130 bytes to at least 401 bytes (or more) to accommodate the proposed credentials sent by the browser. The RADIUS server will also need to update the password verification operation to perform a similar $Verify()$ function.

\subsection{Implementation and Evaluation}\label{Section_implementation}

In this section, we implement the proposed protocol to understand the impact of the proposed change with respect to data and execution overheads. A comparison is made for the password verification use-case for the following three protocols:

\begin{itemize}
    \item Proposed credential protocol in Section \ref{Section_protocol}. For PBKDF2, we used SHA2-512 as the PRF with 1,000 iterations\footnote{Using PBKDF2 with 1,000 PRF iterations is the recommended minimum \cite{moriarty2017pkcs} to make it computationally too expensive for an attacker to carry out brute-force attacks on credentials while not inconveniencing the end-user.}. For digital signing, we used ECDSA $secp256k1$ as the signing algorithm.
    \item Hashed password. Password input by the user is SHA-2 256 hashed at the browser before being sent to the server. The server compares the hashed password with the stored credential which is protected with PBKDF2 (SHA2-512 with 1,000 iterations).
    \item Secure Remote Protocol. This is based on SRP 6 with 2048-bit modulus and SHA2-256 as the hashing algorithm.
\end{itemize}
The implementation is written in NodeJS and run on an Intel Core I5-8250U 8th Gen machine with 8GB RAM. A hardcoded password is used, and all operations are run sequentially where we repeatedly performed a valid front-end to back-end password verification for a single user. 

\subsubsection{Data Overheads} \label{Section_data}
We collected the data usage for transmission and storage for the three protocols in Table \ref{Table_data}. The transmission overhead per authentication for the proposed credential protocol compared to the hashed password is $401-64 = 337$ bytes which is acceptable and is even less than SRP. The storage overhead of $2048 - 1024 = 1024$ bytes per user is also feasible since PBKDF2.

\begin{table}[htpb]
\centering
\caption{Execution benchmark of authentication protocols}
\resizebox{0.8\textwidth}{!}{\begin{tabular}{p{4.0cm}p{2.2cm}p{2.2cm}p{2.2cm}}
\hline
 Parameters  &  Proposed \newline protocol & Hashed \newline Password & SRP \\ \hline
 Data Size in Transmission &  401 bytes & 64 bytes & 640 bytes \\ \hline
 Data Size in Storage & 2048 bytes & 1024 bytes & 64 bytes \\ \hline
 Timing for 1,000 \newline authentications &  33.92 seconds & 12.99 seconds & 23.52 seconds \\ \hline
\end{tabular}
\label{Table_data}}
\end{table}

\subsubsection{Performance} \label{Section_performance}
We measured time taken for each protocol to perform 1,000 successful authentications. Both client and server functions are run natively on the same machine to remove any effects of network latency.

From Table \ref{Table_data}, we observe that using the proposed credential protocol results in the longest execution compared to both hashed password and SRP. Deeper analysis shows that a significant portion of the execution time is due to the server's $Store()$ function where PBKDF2 is used to protect the password and browser public keys. Since we expect the server performance to be deployment-dependent, we can assume that the number of iterations in PBKDF2 can be appropriately chosen to match the required performance. As each authentication only accounts for 33 milliseconds in our test implementation, and can be further optimized through parallel-processing, we decided to keep the proposed protocol as-is.

\subsubsection{Mobile Client}
As a feasibility validation, we tested the client-side execution of the proposed credential protocol on a Samsung Galaxy S10e Android phone running a Qualcomm Snapdragon 855 octa-core. The function was similarly written in NodeJS and run in the Termux (www.termux.com) Android terminal app. The execution of 1,000 $Derive()$ functions took 11.3 seconds which meant that each function took slightly over 11 milliseconds to complete. This confirms our observation that the proposed protocol is computationally feasible.

\section{Conclusion}\label{Section_conclusion}
We have taken a threat analysis approach to identify a design vulnerability within the web password login process. With that understanding, we recommend to deprecate the current \texttt{<input type="password">} input password field within the HTML standard, and propose an input credential field  coupled with a secure protocol that can stop web-based phishing attacks when used in conjunction with anti-virus software. The same proposed protocol supports the use-cases of password enrolment, verification, and change interchangeably and is also suitable for low-entropy password submissions in the case of secondary knowledge-based authentication or out-of-band OTP challenge. We have also analyzed the security of the proposed protocol, deployment issues and performed a data and performance analysis to ascertain the feasibility of the protocol. We believe that while the proposed changes are non-trivial from an implementation point-of-view, the potential benefits in stopping phishing outweigh the costs.

That said, we recognize that the proposed change is not the panacea to solve all password attacks. Social engineering to steal passwords can still happen in other forms such as phone scams, and poor password hygiene will still allow a motivated attacker to compromise the password security of the system. This proposed solution must still be complemented with existing efforts of detection and takedown, plus continued user awareness and training, in order to be effective in removing instances of successful password compromises.

\bibliographystyle{splncs04}
\bibliography{bib}

% % --- Appendix ---%
\appendix

\end{document}